\documentclass{sig-alternate-2013}

\permission{\copyright 2017 International World Wide Web Conference Committee \\ (IW3C2), published under Creative Commons CC BY 4.0 License.}
\conferenceinfo{WWW'17 Companion,}{April 3--7, 2017, Perth, Australia.}
\copyrightetc{ACM \the\acmcopyr}
\crdata{978-1-4503-4914-7/17/04. \\
http://dx.doi.org/10.1145/3041021.3054252 \\
\includegraphics{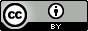}
}

\usepackage{graphicx}
\usepackage{epsfig}
\usepackage{amssymb}
\usepackage{amsmath}
\usepackage{url}

\begin{document}

\title{Towards Understanding the Evolution of the WWW Conference}
%
%
%
%
%

\numberofauthors{3} 
%
\author{
%
%
\alignauthor
Pavel Savov\\
       \affaddr{Polish-Japanese Academy of Information Technology}\\
       \affaddr{ul. Koszykowa 86}\\
       \affaddr{Warsaw, Poland}\\
       \email{pavel.savov@pja.edu.pl}
\alignauthor
Adam Jatowt\\
       \affaddr{Kyoto University}\\
       \affaddr{Yoshida-honmachi, Sakyo-ku}\\       
       \affaddr{Kyoto, 606-8501, Japan}\\
       \email{adam@dl.kuis.kyoto-u.ac.jp}
\alignauthor Radoslaw Nielek\\
      \affaddr{Polish-Japanese Academy of Information Technology}\\
       \affaddr{ul. Koszykowa 86}\\
       \affaddr{Warsaw, Poland}\\
       \email{nielek@pja.edu.pl}
}
\date{7 January 2017}

\maketitle
\begin{abstract}
The World Wide Web conference is a well-established and mature venue with an already long history. Over the years it has been attracting papers reporting many important research achievements centered around the Web. In this work we aim at understanding the evolution of WWW conference series by detecting crucial years and important topics. We propose a simple yet novel approach based on tracking the classification errors of the conference papers according to their predicted publication years. 
\end{abstract}




\section{Introduction}
2017 marks the 26\textsuperscript{th} year of the International World Wide Web (WWW) conference. The conference has served as an important publication venue influencing many researches that center around diverse aspects of the Web. Analyzing its evolution should offer clues as for the characteristics, trends and tendencies in research related to the Web. The objective of this work is then to support analysis of the way in which WWW conference evolved over the years.\\
The evolution of a conference can be analyzed in various ways. In this work, we focus on identifying crucial years in which the key changes occurred. We introduce a novel approach employing a classifier for predicting the publication years of papers. Based on the classification error of papers published at a given year to quantify the importance of that year.\\
The development of research areas \cite{meyer2009} and changes in the topics of academic conferences and journals over time have already been investigated in prior works \cite{saft2014}. Examples of topic model-based approach include  \cite{hall2008} and \cite{mimno2012}. However, to the best of our knowledge, no prior works have focused on the WWW conference series or employed the classification based approach that we propose.

\section{Dataset and Topic Model}
\begin{figure}[h!]
  \centering
  \includegraphics[width=0.4\textwidth]{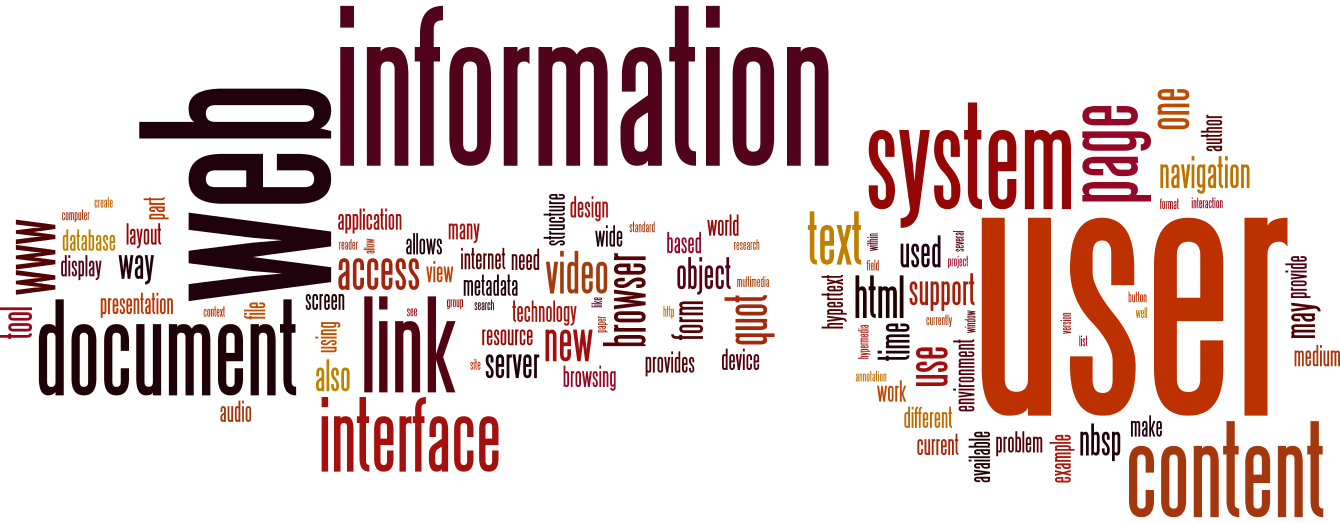}
  \caption{WWW Fundamentals topic}
  \label{fig:topic19}
\end{figure}

\begin{figure}[h!]
  \label{fig:topic16}
  \centering
  \includegraphics[width=0.4\textwidth]{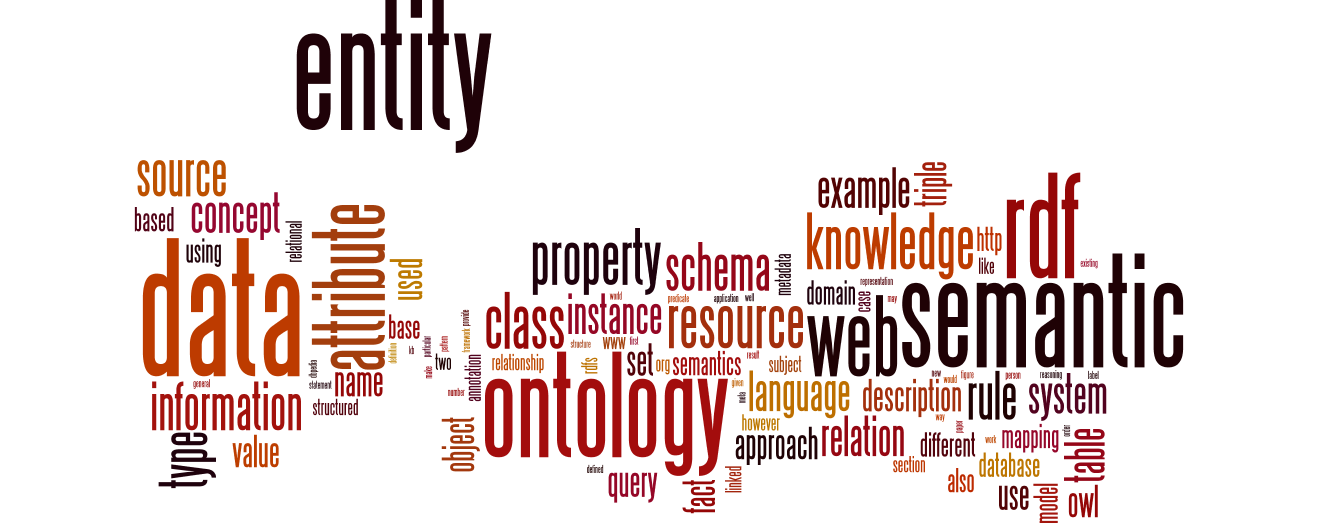}
  \caption{Semantic Web topic}
\end{figure}

Our dataset contains in total 3,105 papers being all full papers and posters published in the proceedings of the WWW conference between the years 1986 and 2016, except 1989, 1994, 1996 and 1997 due to data unavailability. For the purpose of topic extraction and classifier training elements such as page headers and footers, tables, references and acknowledgments have been removed. We have then constructed a 20-topic Latent Dirichlet Allocation \cite{blei2003} model to represent papers by 20-dimensional vectors reflecting their topic distributions.\\
The topic with the highest occurrence probability is the one we call ``WWW Fundamentals'' (Fig. \ref{fig:topic19}). Its popularity however, has been declining for many years in favor of other, more specific topics. This suggests the specialization of research over time. Examples of topics particularly popular in rather short periods include: Semantic Web (Fig. \ref{fig:topic16}), Security, Advertisements, Social Media and Crowdsourcing.

\section{Predicting Publication Years}
We used a multiclass linear SVM classifier trained on the paper set. The classifier has 27 classes -- one for each year when the conference occurred whose papers we managed to collect. Our classifier is available for testing\footnote{\url{http://paper-year-prediction.appspot.com/}}.

\section{Identifying Turnaround Years}
\begin{figure}[h!]
  \centering
  \includegraphics[width=0.5\textwidth]{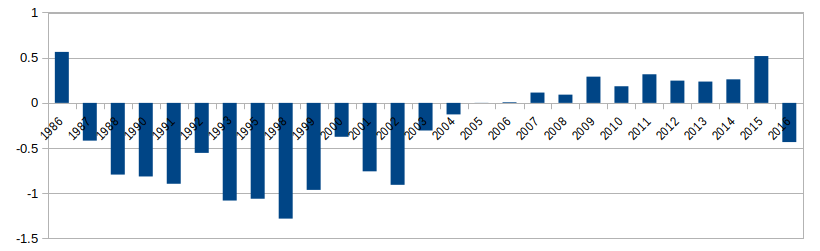}
  \caption{Year importance scores by $S(y)$}
  \label{fig:year-importance-error}
\end{figure}

The underlying motivation behind our approach is as follows. A paper is considered innovative if its topic distribution matches the topic distribution of papers published in the future, especially, in the distant future. The more innovative papers are published in year \textit{y}, the more significant \textit{y} is. In other words -- the greater the mean prediction error for papers in year \textit{y} towards the future, in particular, future distant from \textit{y}, the more important \textit{y} is.\\
Let $Y_b$ be the year of the first conference (i.e. 1986), $Y_e$ -- the year of the last one (i.e. 2016), $P_y$ -- the set of papers published in year $y$ and $\hat{y}(p)$ -- predicted publication year for paper $p$. Next we define $Future_y = \{p \in P_y \mid \hat{y}(p) > y\}$ as the set of documents published in $y$ yet predicted as being ``from the future'', and $Past_y = \{p \in P_y \mid \hat{y}(p) < y\}$ as the set of documents published in $y$ but predicted as ``from the past''.
We can now define the innovation score of year \textit{y} as: 
\begin{equation}
\label{eq:year-importance}
	S(y) = \frac{Err_F(y)}{|P_y|}\cdot N_F(y) - \frac{Err_P(y)}{|P_y|}\cdot N_P(y)
\end{equation}
Where $Err_F(y) = \sum_{p \in Future_{y}}(\hat{y}(p) - y)$ is the total prediction error for all papers in $Future_{y}$ and $Err_P(y) = \sum_{p \in P_{y}}(y - \hat{y}(p))$ is the total prediction error for all papers in $Past_{y}$.
$N_F(y)$ and $N_P(y)$ are the normalization factors for documents predicted as ``from the future'' and ``from the past'' respectively, used to
eliminate bias due to the position of $y$ within [$Y_b$, $Y_e$].
Years with the highest scores are then considered \textit{turnaround years} (see Fig. \ref{fig:year-importance-error}).\\
Note that instead of this classification approach one could try looking into temporal distributions of individual topics for detecting years with many trending topics. The advantage of our method, however, is that it considers all the topic distributions as a whole. It then captures topics that both gain and lose importance as well as the relationships between topics' probabilities in each year.

\section{Results}
\begin{figure}[h!]
  \centering
  \includegraphics[width=0.3\textwidth]{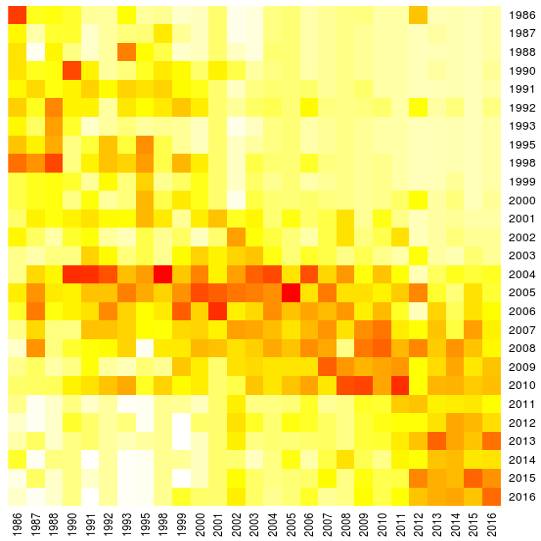}
  \caption{Confusion Matrix}
  \label{fig:confusion-matrix}
\end{figure}

We first define the error function as $E(x) = \hat{y}(x) - y(x)$, where $\hat{y}(x)$ is the predicted year and $y(x)$ is the actual year to measure the prediction quality of our model by \textit{mean absolute error}. The average \textit{mean absolute error} over all folds in a 10-fold cross-validation is 4.27 years.
Fig. \ref{fig:confusion-matrix} shows the confusion matrix as a heat map, where darker shades of red represent higher numbers and paler shades of yellow represent lower numbers. The concentration of higher numbers in two ``squares'' between years 2004 -- 2010 and 2012 -- 2016 indicates that papers are often misclassified within these periods based on their topic distributions. This may suggest trends or ``epochs'' of research. We interpret them as follows:\\
2004 -- 2010: decline of topic ``Web Services'', rising popularity of ``Social Media'', ``Recommendation'', ``Advertisements''\\
2012 -- 2016: rising popularity of ``Crowdsourcing''.
Another important year (see Fig. \ref{fig:year-importance-error}), is 1999, which brings an increase in the popularity of ``Searching'' and ``Data\&Text Mining'' and a decline of ``XML''.

\section{Conclusions}
In this paper we have studied the way in which WWW conference has evolved over the course of its years. We focused in particular on identifying key years that signposted research breakthroughs. For this we have proposed a novel classification approach that predicts publication dates of articles.

\section{Acknowledgments}
This research was supported in part by the Japan's MEXT Grant-in-Aid for Scientific Research (No.15K12158) and by the European Union's Horizon 2020 research and innovation programme under the Marie Sklodowska-Curie (No.690962).

\scriptsize
\bibliographystyle{abbrv}
\bibliography{literature}  
%
%
\balancecolumns
\end{document}